\documentclass{aastex61}

\usepackage{CJK}

\received{\today}
\revised{xxx xx, 2017}
\accepted{xxx xx, 201x}
\submitjournal{ApJL}

\shorttitle{CVSO30c in optical}
\shortauthors{Lee \& Chiang}

\begin{document}
\begin{CJK*}{UTF8}{bkai}
\title{Evidence that the Planetary Candidate CVSO30c is a Background Star from Optical, Seeing-Limited Data}

\correspondingauthor{Chien-Hsiu Lee}
\email{leech@naoj.org}

\author{Chien-Hsiu Lee (李見修)}
\affiliation{Subaru Telescope, NAOJ, 650 North Aohoku Place,
  Hilo, HI 96720, USA}
\author{Po-Shih Chiang (姜博識)}
\affiliation{Graduate Institute of Astronomy, National Central University, No. 300, Jhongda Rd.,
  Jhongli District, Taoyuan City, Taiwan}



\begin{abstract}
  We report serendipitous optical imaging of CVSO30c, an exoplanet candidate associated with the pre-main sequence T Tauri star CVSO30 resides in the 25 Ori stellar cluster. We perform PSF modeling on our seeing limited optical
  image to remove the lights from the host star (CVSO30), allowing us to extract photometry of CVSO30c to be g = 23.2 $\pm$ 0.2 (statistic) $\pm$ 0.1 (systematic) and r = 21.5 $\pm$ 0.1 (statistic) $\pm$ 0.1 (systematic) magnitudes, respectively. This is 170 and 80 times too bright in the g and r-band, respectively, if CVSO30c were an L0 substellar object as suggested by previous studies.
The optical/infrared colors of CVSO 30c are indicative of a stellar, not substellar object, while the object's color-magnitude diagram position is strikingly inconsistent with expected values for a low mass member of 25 Ori. Broad-band photometry for CVSO30c is instead better fit by contaminants such as a background K3 giant or M subdwarf. Our study demonstrates that optical seeing limited data can help clarify the nature of candidate wide separation planet-mass companions in young star-forming regions.
\end{abstract}




\section{Introduction}
\label{sec:intro}
CVSO30 \citep[or PTFO 8-8695][]{2012ApJ...755...42V} is an intriguing candidate
exoplanet system in several aspects. First of all, it resides in the nearby and
young stellar group 25 Orionis \citep{2007ApJ...661.1119B}.
The 25 Ori stellar group has
nearly 200 low-mass PMS (confirmed photometrically and
kinematically) within 1 degree of the bright B star 25 Ori in the Orion OB1a
sub-association located at $\sim$330 pc. With an age of $\sim$ 7 Myr, 25 Ori
stellar cluster if of special interests to exoplanet studies. This is
because pre-main-sequence (PMS) stars, especially with an age of 10 Myr,
play essential roles in our understanding of the formation and evolution of
proto-planetary disks, and planet formation process.

Follow-ups of the 25 Ori group have been
carried out by Young Exoplanet Transit Initiative \citep[YETI,][]{2011AN....332..547N} and
Palomar Transient Factory (PTF) by \cite{2011AJ....142...60V}, leading to the discovery
of an exoplanet candidate CVSO30b. This is a 3-4 Jovian mass exoplanet
candidate transiting a M3 T Tauri star CVSO30 \citep{2012ApJ...755...42V} with a period
of 0.45 days and a separation of 0.008 AU. In addition, high contrast
near-infrared direct imaging with AO also yielded another 4.7 Jovian mass
exoplanet candidate, CVSO30c, associated with the same host star at a separation of
662 AU \citep{2016A&A...593A..75S}.
If both CVSO30b and CVSO30c turned out to be true exoplanets, they can provide
insight on the formation of exoplanets at wide separation. This is because
wide separation exoplanets are
difficult to be reconciled with the core accretion formation mechanism, and
require planet-planet scattering mechanism to explain their wide separation.
One prediction from the planet-planet scattering is that while one exoplanet is
scattered to a wide separation, there will be another exoplanet in the same
system to migrate to very close-in orbits. The existence of a close-in exoplanet
CVSO30b, and a wide separation exoplanet CVSO30c seems to match the planet-planet
scattering scenario.

However, there have been challenges to the exoplanet nature of CVSO30b \citep{2015ApJ...812...48Y}.
Further simultaneous multi-band observations of CVSO30b by \cite{2017PASJ...69L...2O} also contradicted the exoplanet nature, suggesting an origin of circumstellar dust clump instead.
On the other hand, the exoplanet nature of CVSO30c also needs to be examined.
Here we report a serendipitous detection of CVSO30c in the optical band
under seeing-limited conditions. This serendipitous optical imaging contradicts
the exoplanetary candidacy as suggested by \cite{2016A&A...593A..75S}.
Our paper is organized as follows: We describe our observations and data reduction 
in section \ref{sec:obs}. We present our analysis and results, including
several tests on the exoplanet candidacy of CVSO30c using
empirical colors of substellar objects, color magnitude diagram, and
synthetic spectra of substellar objects in section \ref{sec:res}. We
discuss and conclude our results in section \ref{sec:dis}.

\section{Observation and data reduction}
\label{sec:obs}

We performed high cadence time-series photometry monitoring of CVSO30 with
the 3.5m WIYN telescope\footnote{The WIYN Observatory is a joint facility of the
University of Wisconsin-Madison, Indiana University, the National Optical Astronomy Observatory
and the University of Missouri.} at Kitt Peak National Observatory in Arizona, USA. The
observations were carried out
by the One Degree Imager with the 48'$\times$40' field of view and a pixel scale of 0.11''/pixel \footnote{https://www.noao.edu/wiyn/ODI/}. 
The observations were conducted on the nights of February 3rd and 4th, 2017, through the
NASA founded NOAO-WIYN observation program NN-Explore (PI: Lee, ID: 2017-0111).
These data were acquired with the goal of detecting the transit of CVSO 30 b in two filters (g and r-band) to better assess whether or not the object is a planet.
After observations, the data are reduced in a standard manner. The One Degree Imager Pipeline,
Portal, and Archive \citep[ODI-PPA][]{2014SPIE.9152E..0EG} performed bias subtraction,
dark subtraction, flat fielding, and
bad pixel masking. ODI-PPA further aligned each of the individual exposures astrometrically,
and calibrated the photometry using nearby SDSS standard star field observations.

During the photometric monitoring campaign, we noticed traces of a faint
object close to the host star CVSO30 in images with seeing $<$ 0.8'' in both
Sloan g- and r-band images. To further strengthen our detection, we stacked
several exposures using SWarp \citep{2002ASPC..281..228B} to increase the signal-to-noise ratio.
This includes 3 $\times$ 120-sec exposures in the g-band with on-chip seeing
of 0.72-0.73'' and 6 $\times$ 60-sec exposures in the r-band with on-chip seeing
of 0.75-0.79'', respectively. The estimated precision of image registering is 0.03" and 0.04" for g, and r-band, respectively. As the zero-point magnitudes of each individual
exposure is different, we 
calibrate the stacked images photometrically using the
Pan-STARRS Data Release 1\footnote{https://panstarrs.stsci.edu}
\citep{2016arXiv161205560C}. We only used bright (g$<$20 mag and r$<$21 mag), isolated stars for our calibration.
During the photometric calibration, we found
a systematic error at the level of 0.1 mag in both g and r-band.

\begin{figure}
\plotone{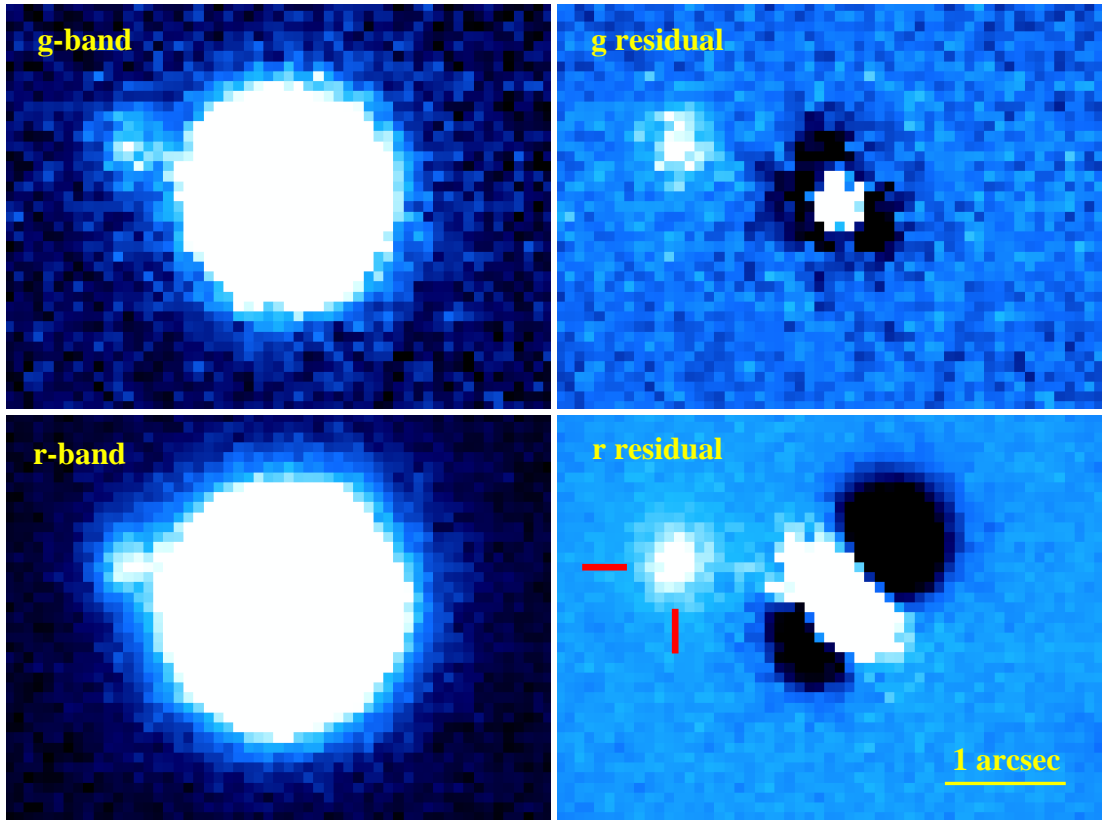}
\caption{Optical imaging of CVSO30c (North is up and East is to the left). Left panel: stacked images, where CVSO30c can
  be seen as a small blob right next to the host star. Right panel: residual images
  after subtracting the host star with PSF modeling. CVSO30c is indicated by
  the red marker.}
\label{fig:f1}
\end{figure}

After stacking, we also model and remove the host star's light with a
point spread function (PSF). The PSF was constructed using bright and isolated
stars in the stacked images using the IRAF routine \textit{psf}. With the
PSF in hand, we then used GALFIT \citep{2002AJ....124..266P,2010AJ....139.2097P}
to subtract the host star’s flux, yielding
to a clear detection of this faint object (as shown in Fig. \ref{fig:f1}).
We then obtained optical
photometry of this faint object from the stacked images using SExtractor
\citep{1996A&AS..117..393B}, deriving
g = 23.2 $\pm$ 0.2 (statistic) $\pm$ 0.1 (systematic) and
r = 21.5 $\pm$ 0.1 (statistic) $\pm$ 0.1 (systematic) 
magnitudes.
We note that \cite{2016A&A...593A..75S} also carried out z-band
observations using lucky image technique with AstraLux onboard the Calar
Alto 2.2m telescope. While they did not detect CVSO30c with AstraLux, they
were able to constrain an upper limit of 20.5 mag in z-band.
This is consistent with our faint detection in the g- and r-bands.

Finally, we measured the separation of this faint object to the star
CVSO30 to be 1.83''$\pm$0.04'', with a position angle of 71.6$\pm$0.8
degrees (East from North). We note that the imager has a north position angle uncertainty of $\sim$ 0.26 degrees.
The separation and positional angle of this object in the optical are in good agreement with the separation
of $\sim$ 1.85'' and the position angle of $\sim$ 70 degrees
from \cite{2016A&A...593A..75S},
hence confirmed this optically faint object is indeed CVSO30c. 

\section{Results}
\label{sec:res}
With the optical photometry, we can test whether the exoplanetary
candidacy of CVSO30c holds.
First of all, \cite{2016A&A...593A..75S} reported that this object is a young and substellar, with a spectral type later than L0.
  For comparison, we check other known substellar objects with similar spectral type and age.
  ROXs 42B, for example, has an effective temperature of $\sim$ 2,000 K,
  a spectral type of L0$\pm$1, and a low gravity due to its youth. Another case is $\beta$ Pic b,
  which has an effective temperature of 1,600-1,700 K, a spectral type of L3 or later, and also a low gravity due to its youth.
  In addition, young, low gravity L and L/T transition objects have redder colors than their field counterparts.
Therefore, the comparison to L0 is a best-case scenario for CVSO30c being a planet.
The intrinsic optical and
infrared colors of objects as late as L0 are well well known and described in
\cite{2007AJ....134.2340K}. The predicted g-J and r-J colors are 9.2 and 6.7
magnitudes. According to \cite{2016A&A...593A..75S}, CVSO30c has J=19.6 mag;
assuming the object is an L0 substellar object, it should be fainter than g=28.8
and r=26.3 magnitudes. However, given our measurement of CVSO30c,
it is 180 (in g) and 70 (in r) times brighter than an L0 objects,
  as suggested by \cite{2016A&A...593A..75S}.

Secondly, we can also check whether CVSO30c resides in the 25 Ori stellar cluster
  using color magnitude diagram, by combing the optical photometry presented here and
  the infrared photometry in \cite{2016A&A...593A..75S}.
\cite{2014MNRAS.444.1793D} have identified 77 low mass members of the 25 Ori stellar
cluster spectroscopically. Their members all follow a well-defined 7 Myr isochrone
from Baraffe et al. (1998) on the I vs. I-J color magnitude diagram.
We cross matched the catalogue by \cite{2014MNRAS.444.1793D} with the PS1 DR1 catalog \citep{2016arXiv161205560C}
to retrieve g and r magnitudes of the cluster members.

\begin{figure*}
  \centering
  \includegraphics[scale=1.2]{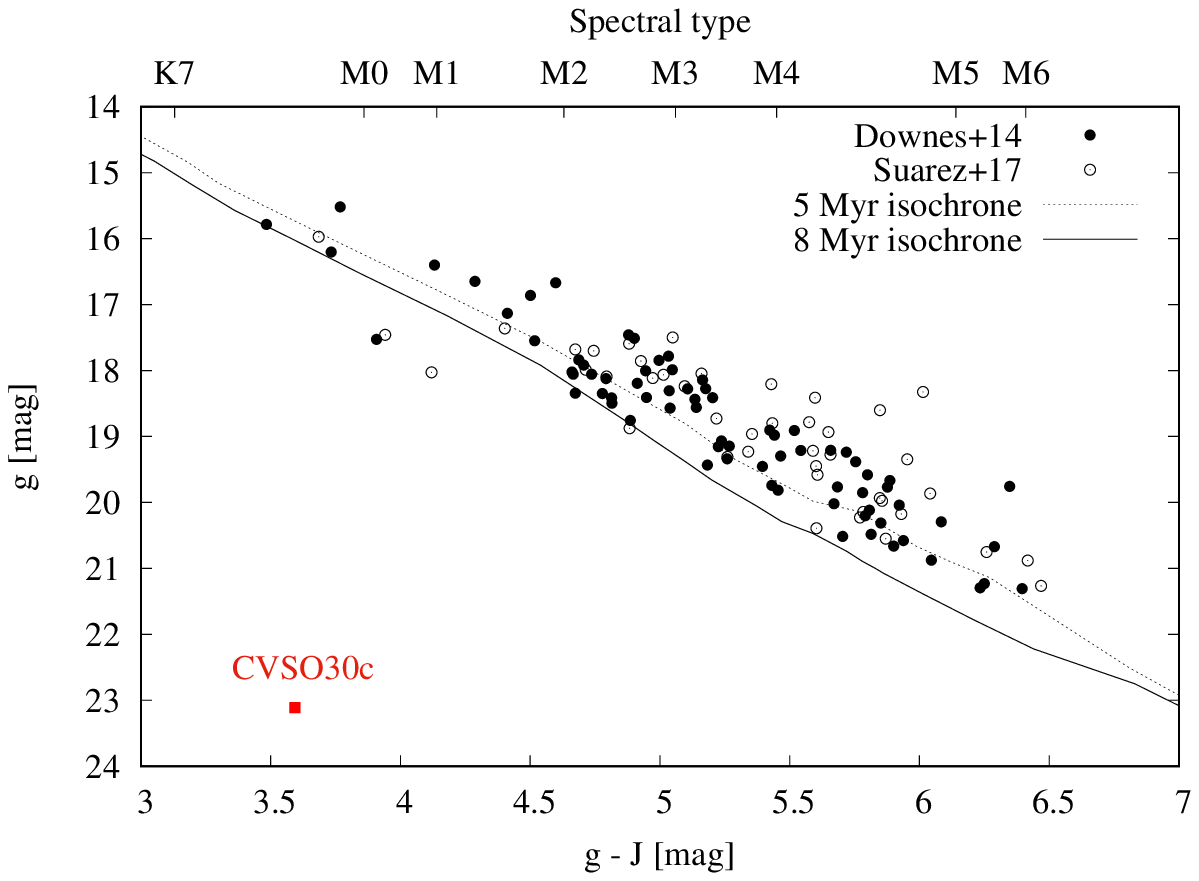}
          \includegraphics[scale=1.2]{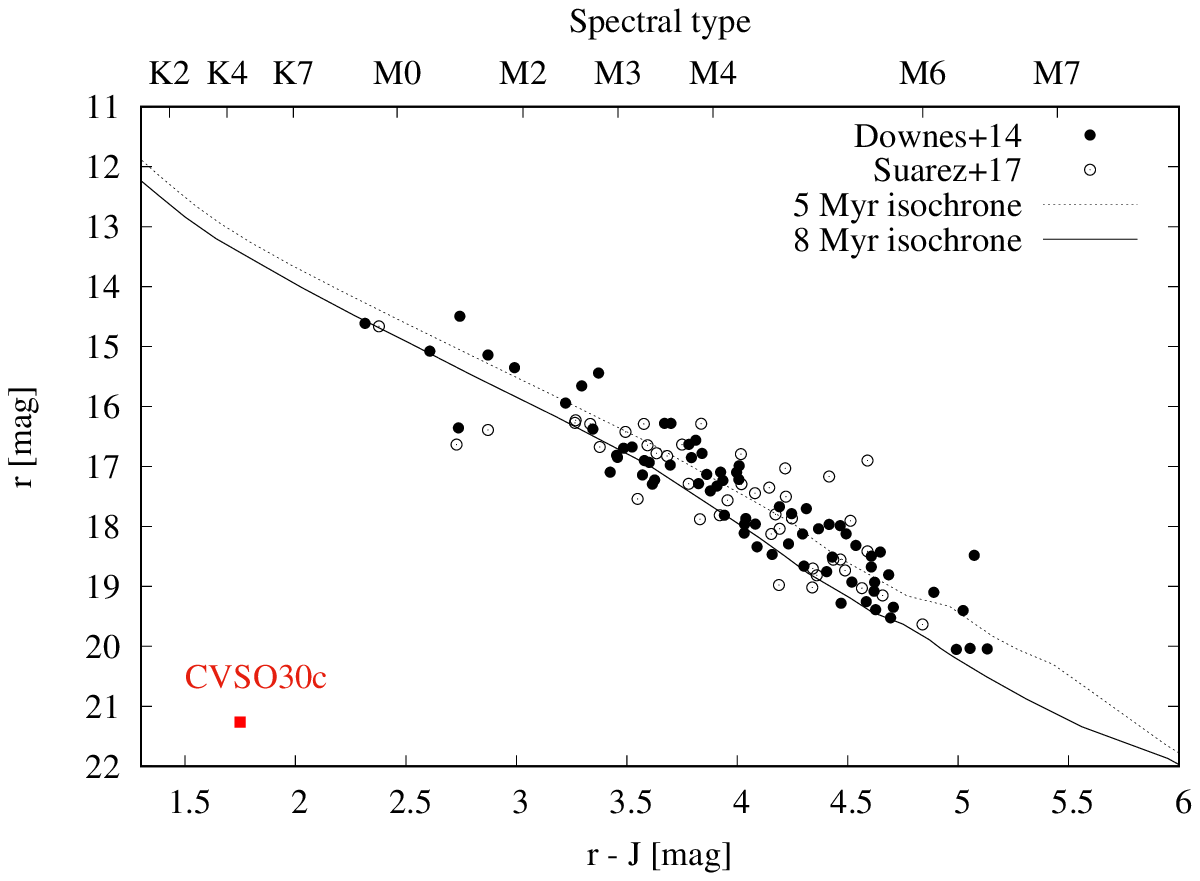}
          \caption{g vs. g-J and r vs. r-J color magnitude diagram of the 25 Ori stellar group. We plot the spectroscopically confirmed
            25 Ori stellar group low mass members from \cite{2014MNRAS.444.1793D} and
            \cite{2017AJ....154...14S}, marked in solid and empty circles. We also plot the isochrone by \cite{2015A&A...577A..42B}
            assuming a distance modulus of 7.6 magnitudes, with ages of 8 Myr (solid line) and 5 Myr (dashed line), respectively. 
            CVSO30c (marked in red), on the contrary to the spectroscopically confirmed low mass members of 25 Ori stellar group, does not follow
            the isochrones. This suggests that CVSO30c does not belong to the 25 Ori stellar group.}
\label{fig:f2}
\end{figure*}

In addition to \cite{2014MNRAS.444.1793D},
\cite{2017AJ....154...14S} also reported 50 low mass members (including their
positions and SDSS photometry) in the 25 Ori stellar group and
Orion OB1a sub-association, which are all co-distant at 338$\pm$66 pc.
Color-magnitude diagrams clarify whether CVSO30c is consistent with being a low-mass member of 25 Ori. As shown in Fig. \ref{fig:f2},
  CVSO30c is strikingly inconsistent with the color-magnitude diagram position expected for a low-mass member of 25 Ori.

\begin{figure*}
  \centering
  \includegraphics[scale=1.2]{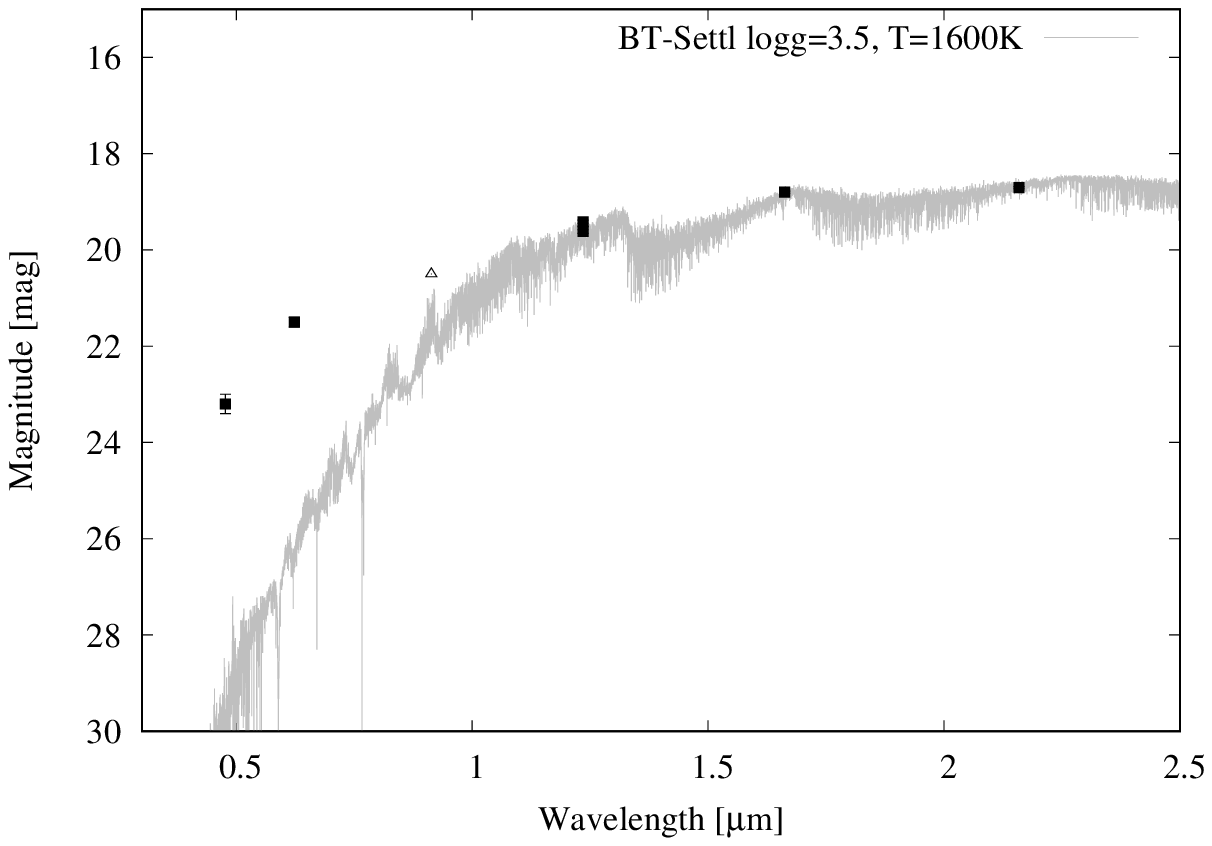}
  \caption{Spectral energy distribution of CVSO30c. We show the optical photometry from this work and the infrared photometry from \cite{2016A&A...593A..75S}, both marked in squares. We also show the z'-band upper limit by \cite{2016A&A...593A..75S}, marked in open triangle. We compare the broad-band photometry to a synthetic spectrum of a slightly reddened L0 substellar object with log$g$ = 3.5,  effective temperature of 1600 K, and Av=0.3, as suggested by \cite{2016A&A...593A..75S}. The synthetic spectrum (shown in gray) is retrieved from the Phoenix Web Simulator, adopting the BT-Settl model \citep{2012RSPTA.370.2765A}. For comparison, we scaled the synthetic spectrum to the infrared photometry of CVSO30c. While the infrared photometry is in good agreement with the synthetic spectrum of a substellar object, the optical photometry is in tension with this synthetic spectrum by 5 magnitudes brighter in both g and r-band.}
\label{fig:f3}
\end{figure*}

Further more, we can examine whether the
SED of CVSO30c
fits an L0 substellar
object as suggested by \cite{2016A&A...593A..75S}.
From the Phoenix Web Simulator\footnote{https://phoenix.ens-lyon.fr/simulator/index.faces},
  we retrieved the synthetic spectrum of an object with log$g$ = 3.5 and effective temperature of 1,600 K, as suggested
  by \cite{2016A&A...593A..75S}.
For comparison,
we scaled the synthetic spectrum to match the infrared photometry, as shown in
Fig. \ref{fig:f3}. The results indicate that while we can match the synthetic spectrum
to the infrared photometry, the optical photometry does not follow the synthetic spectrum,
and is off by 5 magnitudes in both the g and r-band.
This suggests that the optical photometry of CVSO30c is in tension
with the exoplanetary scenario.

\section{Discussion and conclusion}
\label{sec:dis}
To summarize, our serendipitous optical detection and analysis of CVSO30c provides evidence against the object being
  a directly-imaged planetary companion as proposed by \cite{2016A&A...593A..75S}. Its optical/infrared color disagrees with values
  expected for an L0 or later substellar object from the SED library of \cite{2007AJ....134.2340K}. Furthermore, the optical/infrared
  color-magnitude diagram position of CVSO30c is strongly inconsistent with the locus of spectroscopically confirmed low mass 25 Ori
  members and with theoretical isochrones appropriate for young stars and substellar objects. Finally, CVSO30c's broadband photometry
  strongly disagrees with the synthetic spectrum of a low gravity L0 model atmosphere from the BT-Settl grid.

\begin{figure*}
  \centering
  \includegraphics[scale=1.2]{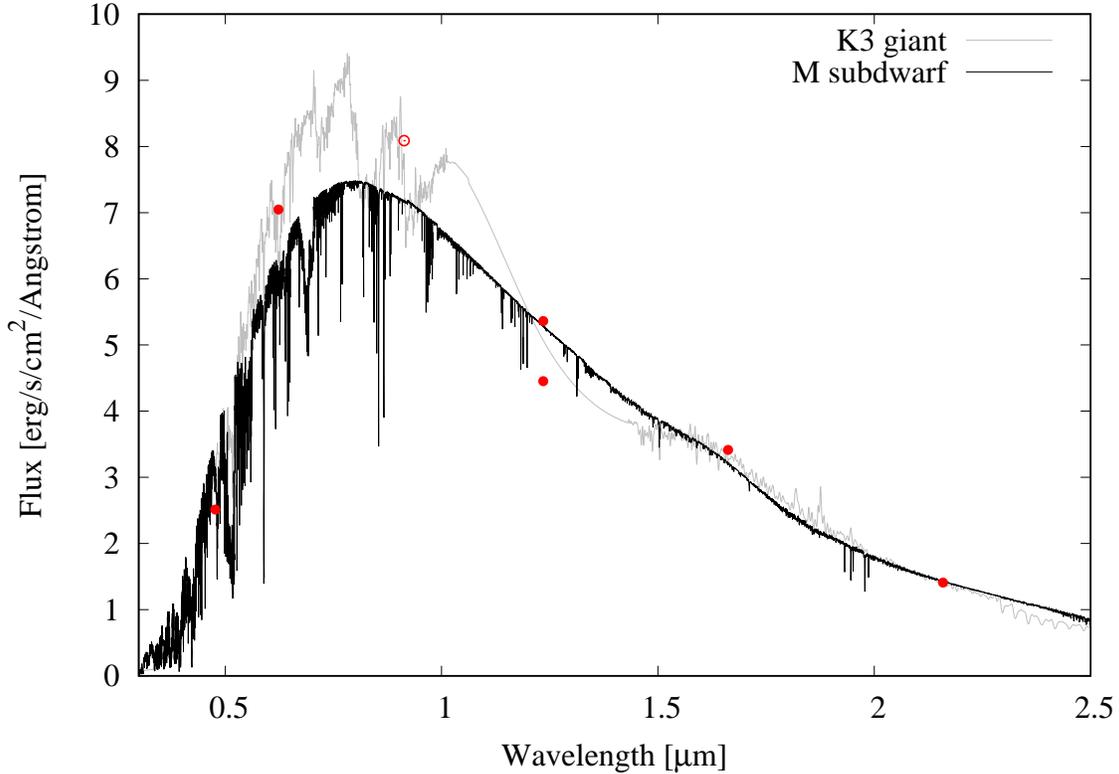}
  \caption{Comparison of CVSO30c broad-band photometry with an K-giant and a M subdwarf.  The grJHK photometry are marked in red solid circles, while the z-band upper limit by \cite{2016A&A...593A..75S} is marked in red open circle. We show the synthetic spectra of a K3 Giant from the spectral library of \cite{1998PASP..110..863P}, with an Av=0.6, as well as a synthetic spectra of a M subdwarf with T$_{eff}$=3,800K, log$g$=5.5, Z=-2.0 from BT-Settl \citep{2012RSPTA.370.2765A}, with an Av=0.26. Both provide a good fit to the CVSO30c broad-band photometry, suggesting that it is a background star.}
\label{fig:f4}
\end{figure*}

Given the optical detection, it is intriguing to understand the
  inconsistency between our analysis and that from \cite{2016A&A...593A..75S}.
  First of all, \cite{2016A&A...593A..75S} argued that the z' non-detection ruled out objects hotter
  than 3,300 K. However, consulting to the intrinsic colors of stellar objects compiled by
  \cite{2007AJ....134.2340K}, we can see that the optical and infrared colors by
  \cite{2016A&A...593A..75S} can only rule out objects with spectral type of K2 or earlier,
  i.e. T$_{eff} > $ 5,000 K. This then allows objects with effective temperatures of 3,300 to 5,000 K as
  possible contaminants, which are not considered in \cite{2016A&A...593A..75S}. Yet from
  our r-J colors the CVSO30c seems reasonable well matched to a K5-M0 star (as shown in Fig. \ref{fig:f2}.
  Secondly, while \cite{2016A&A...593A..75S} argued that their IR data analysis rules out a background star,
  they did not fully explore the range of background objects. For example, the giants they considered are all later than M4.5.
  In fact, when we compare the broad-band photometry with a wide variety of spectral types, we found that CVSO30c is more likely
  to be a reddened K3 super giant or a metal-poor early-M subdwarf (Fig. \ref{fig:f4}, which were not considered by \cite{2016A&A...593A..75S}.

  We note that it might be possible that our optical detection is not associated with CVSO30c, i.e. a background star that
  is not detected at longer wavelengths in JHK.
However, consulting the SED library of \cite{2007AJ....134.2340K}, the g-r colors alone suggest an object with an M2-M3 spectral type - not an L0 or later object - which correspondingly have r-J and r-K colors of 3.03-3.46, 3.9-4.34, respectively. Such and object, with an apparent magnitude of J=18-18.5 and K=17.2-17.6 at the distance of 25 Ori (without extinction correction), would have been comparable in brightness to CVSO30c or brighter and detectable in the \cite{2016A&A...593A..75S} near-IR data. Combined with the consistent spatial alignment between our optical detection and the infrared, we conclude that our optical detection is the same object seen by \cite{2016A&A...593A..75S} and labeled CVSO30c.

This work demonstrates that optical imaging under good seeing-limited conditions can clarify the nature of candidate,
  very wide-separation companions to young stars in star-forming regions. Given the large implied physical separations of these objects
  (hundreds of au), proper motion analyses are poorly suited to identify background stars. However, the optical/infrared colors of
  young planetary-mass objects significantly differ from most contaminating background sources. Such observations are also efficient.
  In this case, optical imaging using a 3.5-meter telescope with a 2 minute integration time in g-band or 1 minute in r-band can already
  rule out a candidate planetary companion. Follow-up spectroscopic observations can then focus on candidates whose optical data do not
  reveal them to be background objects.

  \acknowledgments
We are indebted to the referee, whose insightful comments and suggestions improved this manuscript significantly.

Based on observations at Kitt Peak National Observatory, National Optical Astronomy
Observatory (NOAO Prop. ID: 2017A-0111; PI: Lee), which is operated by the Association
of Universities for Research in Astronomy (AURA) under a cooperative agreement with the National Science Foundation. 

Data presented herein were obtained at the WIYN Observatory from telescope time allocated to NN-EXPLORE through
the scientific partnership of the National Aeronautics and Space Administration, the National Science Foundation,
and the National Optical Astronomy Observatory.

%

\vspace{5mm}
\facilities{WIYN(ODI)}

\end{CJK*}

\end{document}